\documentstyle[twoside,fleqn,espcrc2,epsf]{article}

\newcommand{\AmS}{{\protect\the\textfont2
  A\kern-.1667em\lower.5ex\hbox{M}\kern-.125emS}}

\title{Lattice Analysis of Two-Point Hadronic Correlators
   in the QCD Vacuum}

\author{M.-C. Chu,\address{Kellogg Laboratory, California Institute of
        Technology, 106-38, \\
        Pasadena, California 91125  U.~S.~A.}
        J.~M.~Grandy,\address{T-8 Group, MS B-285, Los Alamos National
        Laboratory, \\
        Los Alamos, New Mexico 87545 U.~S.~A.}
        S.~Huang,\address{FM-15, Department of Physics,
        University of Washington, \\
        Seattle, Washington 98195 U.~S.~A.}
        and
        J.~W.~Negele\address{Center for Theoretical Physics,
        Laboratory for Nuclear Science, and Department of Physics, \\
        Massachusetts Institute of Technology, Cambridge,
        Massachusetts 02139 U.~S.~A. }
        \thanks{This work is supported in part by funds provided by
the U.~S. Department of Energy (D.~O.~E.) under contracts
\#~DE-AC02-76ER03069 and \#~DE-FG06-88ER40427, and the National Science
Foundation under grant \#~PHY~88-17296.}}

\begin{document}

\begin{abstract}
     Results from the first lattice QCD analysis of vacuum correlators
of local hadronic currents using dispersion relations are presented.
We have explored the vector, pseudoscalar, axial, and scalar meson
channels, and the proton-like and delta-like baryon channels.  The
lattice results are shown to agree qualitatively with experimental
results in channels where experimental data exist, and shed insight
into interacting instanton approximations and sum rule calculations in
the other channels.
\end{abstract}

\maketitle

\section{INTRODUCTION}
Two point correlators of local interpolating hadronic currents are a
useful means of studying the QCD vacuum.  These correlators have been
measured experimentally for the vector and axial channels, and have
been computed phenomenologically using interacting instanton
approximations\cite{Shuryak89} and QCD sum rules\cite{Ioffe,FZ81} in
the remaining channels.  The current experimental and phenomenological
predictions of the correlators have been reviewed extensively by
Shuryak\cite{Shuryak92}.  Lattice calculations provide a means of
investigating the correlators from the first principles of QCD,
although at present the lattice work is in an exploratory stage.  We
therefore emphasize the qualitative features of the comparison between
our lattice results and the predictions and demonstrate the ability of
lattice calculations to refine sum rule calculations.

\begin{table*}[hbt]
\setlength{\tabcolsep}{1.08pc}
\renewcommand{\arraystretch}{1.3}
\newlength{\digitwidth} \settowidth{\digitwidth}{\rm 0}
\catcode`?=\active \def?{\kern\digitwidth}
\caption{Hadronic Current Correlation Functions}
\label{tab:Currents}
\begin{tabular}{lrrr}
\hline
                   \multicolumn{1}{l}{Channel}
                 & \multicolumn{1}{r}{Current}
                 & \multicolumn{1}{r}{Correlator}
                 & \multicolumn{1}{r}{$f_p(s)$}      \\
\hline
Vector       & $J^V_\mu = \bar{u}\gamma_\mu d $
        & $\left\langle 0|T\left[ J^V_\mu(x) \bar{J}^V_\mu(0)
                  \right] |0\right\rangle $
        & $ {{1}\over{12\pi^2}}$ \\
Axial        & $J_\mu^A = \bar{u}\gamma_\mu \gamma_5 d $
        & $\left\langle 0|T\left[ J_\mu^A(x) \bar{J}_\mu^A(0)
                  \right]|0\right\rangle $
        & $ {{1}\over{12\pi^2}}$ \\
Pseudoscalar & $J^P = \bar{u}\gamma_5 d $
        & $\left\langle 0|T\left[ J^P(x) \bar{J}^P(0)
                  \right]|0\right\rangle $
        & $ {{3s}\over{8\pi^2}}$ \\
Scalar       & $J^S = \bar{u} d $
        & $\left\langle 0|T\left[ J^S(x) \bar{J}^S(0)
                  \right]|0\right\rangle $
        & $ {{3s}\over{8\pi^2}}$ \\
Nucleon      & $J^N = \epsilon_{abc}(u^a C\gamma_\mu u^b) \gamma_\mu
                  \gamma_5 d^c $
        & ${{1}\over{4}}\,{\rm Tr}\, \left(\left\langle 0|
                  T\left[ J^N(x) \bar{J}^N(0)\right]|0\right\rangle
                  x_\nu \gamma^\nu \right)$
        & $ {{s^2}\over{64\pi^4}}$ \\
Delta        & $J^\Delta_\mu = \epsilon_{abc} (u^a C\gamma_\mu u^b) u^c $
        & ${{1}\over{4}}\,{\rm Tr}\, \left(\left\langle 0|
                  T\left[ J^\Delta_\mu(x) \bar{J}^\Delta_\mu(0) \right]|0
                  \right\rangle x_\nu \gamma^\nu \right)$
        & $ {{3s^2}\over{256\pi^4}}$ \\
\hline
\end{tabular}
\end{table*}

The hadronic currents we use and their correlators are listed in Table
1.  The extra $(x_\nu \gamma^\nu)$ factor for the baryons selects the
part which is stable in the chiral limit.  In the next section we
briefly review the existing predictions of the correlators.  Section 3
describes the method of our lattice calculations.  Finally, we present
our lattice results and describe the comparisons with predictions.

\section{CORRELATOR PREDICTIONS}

The correlator in the vector channel is determined directly from
hadron production in $e^+e^-$ annihilation experiments.  The axial
correlator is also known experimentally from the isospin-violating
$\tau \rightarrow 3\pi$ reaction, although the direct experimental
determination is limited to $Q^2 < 1.5\,{\rm GeV}^2$ due to the mass
of the $\tau$.  In the pseudoscalar channel the correlator is fit
phenomenologically using the dispersion form described below.  For the
scalar, the experimental data is sketchy, so the best means of
predicting the correlator is by using the interacting instanton
approximation (AII)\cite{Shuryak89}.  For the baryons, the threshold
and current coupling parameters obtained from sum rule
calculations\cite{Ioffe,FZ81} are entered into the dispersion
relations to obtain predicted correlations.

The form of the correlator in momentum space, $\hat R(q) = \int d^4x\,
e^{iqx} R(x)$, is given by the standard dispersion relation\cite{ours}
\begin{equation}
 \hat R(q)
  = \left\{\matrix{ 1\cr 3q^2\cr-iq^\mu\gamma_\mu\cr}\right\}
\left( \int ds {f(s)\over s-q^2} + \ldots \right)
\end{equation}
where contact terms and terms with other Dirac structures have been
omitted.  The top term in the brackets is used for the pseudoscalar
and scalar channels, the middle term for the vector and axial, and the
bottom term for the baryons.  The physical spectral density $f(s)$ is
phenomenologically approximated using a resonance contribution from
the ground state $f_r(s) =
\lambda^2\delta(s-M^2)$ where $M$ is the resonance mass and $\lambda$
is the current coupling to the resonance state, and a contribution
from a continuum of excited states above the threshold energy $s_0$,
$f_c(s) = f_p(s) \theta(s-s_0)$.  We use asymptotic freedom to
approximate the perturbative correlators with the corresponding free
correlators in Table 1.  In position space the correlators are given
by $R(x) = R_r(x) + R_c(x)$ where
\begin{equation}
R_r(x)
  = \lambda^2 M g_r(x) K_p(Mx)
\end{equation}
\begin{equation}
R_c(x) = \int^{\infty}_{s_0} ds\ f_p(s)\, g_c(s,x) K_p(\sqrt{s}\,x) \, .
\end{equation}
The functions $g_r$ and $g_c$ are listed in Table 2.  The Bessel
function orders are $p=1$ for mesons and $p=2$ for baryons.  This form
for $R(x)$ is the basis for phenomenological fits by
Shuryak\cite{Shuryak92} and also our fits to lattice data.

\begin{table}[hbt]
\setlength{\tabcolsep}{1.35pc}
\renewcommand{\arraystretch}{1.3}
\settowidth{\digitwidth}{\rm 0}
\catcode`?=\active \def?{\kern\digitwidth}
\caption{Correlator Fitting Functions}
\label{tab:fitfns}
\begin{tabular}{lrr}
\hline
                   \multicolumn{1}{l}{Channel}
                 & \multicolumn{1}{r}{$g_r(x)$}
                 & \multicolumn{1}{r}{$g_c(s,x)$}         \\
\hline
V,A    & $3M^2x^{-1}$       & $3s^{3/2} x^{-1}$  \\
P,S    & $x^{-1}$           &  $s^{1/2} x^{-1}$  \\
N,$\Delta$ & $M$            &  $s$       \\
\hline
\end{tabular}
\end{table}

\section{LATTICE CORRELATORS}
\subsection{Background}
We compute lattice correlators on $16$ independent $16^3\times 24$
configurations at $\beta=5.7$, with a lattice spacing normalized by
the proton mass of $a=.168\, {\rm fm}$.  Hard wall boundary conditions
are imposed at the end time slices with periodic boundary conditions
in the three spatial directions.  Propagators, generated using a
localized source in the central time slice, have been previously
computed by Soni {\it et al.\/}\cite{Soni}.  Free particle
correlators, $R_0(x)$, are calculated analytically on a very large
lattice, $(48)^4$, to eliminate boundary effects at separations below
$2\,$ fm.  In practice, we compare $R(x)/R_{0}(x)$ on the lattice with
predictions of the same, in order to more closely observe the onset of
asymptotic freedom and the effects of quark interactions.

\subsection{Lattice Artifacts}

\begin{figure}[htb]
\epsfysize=3.0in
\epsfxsize=2.9in
\epsfbox{pi.ps}
\caption{Measured pseudoscalar vacuum correlators for five lattice
quark masses and the extrapolation to the physical pion mass.  Solid
curves are fits through lattice results using Eqs.~2-3, and dot-dash
curve is the phenomenological result calculated by
Shuryak[4].}
\label{fig:pionplot}
\end{figure}
It is necessary to account for lattice artifacts when extracting
physical results from the computed quantities.  First, periodic
boundary conditions introduce leakage between Brillouin zones
surrounding the point sources.  This leakage is corrected by
subtracting contributions from image sources from the lattice data,
and self-consistently fitting a curve through the corrected points.
The Cartesian lattice introduces artificial directional anisotropy.
To counteract this we select only points near the body diagonal, $\hat
d= {{1}\over{\sqrt{3}}} (1,1,1)$, so that $\hat x \cdot \hat d \ge
0.9$.  In this direction, the free propagator agrees closely with the
continuum result, and we thus believe that the interacting results are
most reliable in this direction.  In our current phase we use
separations $x$ within the central time slice to avoid contamination
from the hard wall, and we plan in the future to include a few slices
from the center.  For the purposes of fitting, we consider $R(x)$ as a
function of the magnitude of the separation and combine lattice
separations in one-lattice-unit bins.  It is desirable to normalize
the ratio $R(x)/R_{0}(x)$ to unity at an infinitesimal separation
but we actually normalize at our smallest separation, $\sqrt{3}a =
0.29\,{\rm fm}$.

\begin{figure}[t]
\epsfysize=3.8in
\epsfxsize=2.9in
\epsfbox{va.ps}
\caption{Vacuum correlators extrapolated to physical pion mass for
meson channels.  Solid curve is dispersion fit through lattice data,
dashed curves are results derived from experimental
data[4], and dotted curves are AII
calculation[1].}
\label{fig:mesonplot}
\end{figure}
We also must contend with the inability to compute propagators at the
physical pion mass.  We compute the correlators at quark masses of
$m_q =\,$ 351, 199, 110, 67, and 25 MeV and extrapolate the
correlators at the lightest four quark masses using a best quadratic
fit to the quark mass $m_q=8\,{\rm MeV}$ which corresponds to the
physical pion mass of $140 \,{\rm MeV}$.  In the pseudoscalar channel
which becomes infinite in the chiral limit the logarithm of the
correlator is extrapolated.  As an example, the extrapolation of the
pseudoscalar is explicitly plotted (Fig.~1).  In the other channels
only the extrapolated results are shown.

\begin{figure}[t]
\epsfysize=3.8in
\epsfxsize=2.9in
\epsfbox{nd.ps}
\caption{Vacuum correlators extrapolated to physical pion mass for
baryon channels.  Solid curve is dispersion fit through lattice data,
compared with results based on sum rule calculations by Ioffe (dashes)
and Farrar {\it et al.\/} (dot-dash).}
\label{fig:baryonplot}
\end{figure}

\section{RESULTS}

In the vector channel, it is remarkable\cite{Shuryak92} that the ratio
$R(x) \over R_{0}(x)$ remains close to 1 over the range of separations
plotted, although $R(x)$ falls by several orders of magnitude.  The
lattice calculation (Fig.~2a) is consistent with this result, and has
similar features to the experimental result.  In the axial channel
(Fig.~2b) the lattice data are fitted with the continuum distribution
$R_c(x)$ only since we cannot resolve a resonance contribution from
the continuum with our data.  Again, the lattice result is
qualitatively similar to the phenomenology but the rising tail, due to
pion mixing, is difficult to reproduce.  In the pseudoscalar channel
the extrapolation is shown explicitly, and the extrapolated result at
the physical pion mass is close to Shuryak's fit.  The scalar
channel, not plotted here, is difficult to extrapolate and subject to
large statistical errors.  As in the axial channel we see no clear
resonance contribution for the scalar.

There are no direct experimental determinations of the baryon
correlators, so we compare in figure 3 the lattice results with the
dispersion fits based on sum rule calculations by Farrar {\it et al.}
\cite{FZ81} (dotdash) and Ioffe\cite{Ioffe} (dashes).  The good
dispersion fits to the baryon data and the wide disparity between the
sum rule calculations suggest that future lattice results can be used
to refine sum rule calculations.

This exploratory calculation of two-point hadronic correlators
produces qualitative agreement with available experimental results and
phenomenological fits in most channels.  This agreement motivates
further, improved lattice calculations which can be used in tandem
with previously established methods to understand the behavior of
hadronic correlators in the QCD vacuum.


\begin{thebibliography}{9}
\bibitem{Shuryak89} E.~Shuryak, {\it Nucl. Phys.\/} {\bf B328}, 102
(1989).

\bibitem{Ioffe} B.~L.~Ioffe, {\it Nucl. Phys.\/} {\bf B188}, 317 (1981);
        V.~M.~Belyaev and B.~L.~Ioffe, {\it Sov. Phys. JETP\/} {\bf 83},
        976 (1982).
\bibitem{FZ81} G.~Farrar, H.~Zhoang, A.~A.~Ogloblin and
        I.~R.~Zhitnitsky, {\it Nucl. Phys.\/} {\bf B311}, 585 (1981).
\bibitem{Shuryak92}E.~Shuryak, Stony Brook preprint SUNY-NTG-91/45,
to appear in {\it Rev. Mod. Phys.\/} (1992).
\bibitem{ours} M.-C.~Chu, J.~M.~Grandy, S.~Huang, and J.~W.~Negele,
        MIT Preprint CTP\#2113 (1992) (hep-lat 9208030).
\bibitem{Soni} A.~Soni, {\it National Energy Research Supercomputer Center
Buffer} {\bf 14}, 23 (1990).  Relevant details are presented in C.~Bernard,
T.~Draper, G.~Hockney and A.~Soni, {\it Phys. Rev.\/} {\bf D38}, 3540 (1988).

\end{thebibliography}
\end{document}